\documentclass[pra,twocolumn,amssymb,amsmath]{revtex4-1}
\usepackage{CJK}
\usepackage{graphicx}
\usepackage{epsfig,color}
\usepackage{amsmath,amssymb,eufrak}
\usepackage{amsthm}
\usepackage{enumerate}
\usepackage{hyperref}
\usepackage{color,amsxtra,amsfonts,bm}

\newcommand{\be}{\begin{equation}}
\newcommand{\ee}{\end{equation}}
\newcommand{\bea}{\begin{eqnarray}}
\newcommand{\eea}{\end{eqnarray}}
\begin{document}

\title{Removing Staggered Fermionic Matter in $U(N)$ and $SU(N)$ Lattice Gauge Theories}

\date{\today}

\author{Erez Zohar}
\address{Max-Planck-Institut f\"ur Quantenoptik, Hans-Kopfermann-Stra\ss e 1, 85748 Garching, Germany,}
\address{Munich Center for Quantum Science and Technology (MCQST), Schellingstr. 4, D-80799 M\"unchen, Germany,}
\address{Racah Institute of Physics, The Hebrew University of Jerusalem 91904, Givat Ram, Jerusalem, Israel.}

\author{J. Ignacio Cirac}
\address{Max-Planck-Institut f\"ur Quantenoptik, Hans-Kopfermann-Stra\ss e 1, 85748 Garching, Germany,}
\address{Munich Center for Quantum Science and Technology (MCQST), Schellingstr. 4, D-80799 M\"unchen, Germany,}

\begin{abstract}
Gauge theories, through the local symmetry which is in their core, exhibit many local constraints, that must be taken care of and addressed in any calculation. In the Hamiltonian picture this is phrased through the Gauss laws, local constraints that restrict the physical Hilbert space and relate the matter and gauge degrees of freedom. In this work, we present a way that uses all the Gauss laws in lattice gauge theories with staggered fermions for completely removing the matter degrees of freedom, at the cost of locally extending the interaction range, breaking the symmetry and introducing new local constraints, due to the finiteness of the original local matter spaces.
\end{abstract}

\maketitle

\section{Introduction}

The principle of gauge invariance is very important in physics, as it describes the fundamental forces and interactions in the standard model of particle physics. The excitations of gauge fields - gauge bosons - mediate local interactions between the matter particles. It is a local symmetry, that involves many constants of motion defined at each point of space, giving rise to local constraints (Gauss laws) satisfied by both the matter and gauge degrees of freedom.

Gauge symmetry also implies a redundancy in the mathematical description. Only gauge invariant quantities - combinations of the fields that are invariant under the gauge transformations - are considered physical. This immediately raises the question whether this redundancy could be lifted, at least partially, by simply solving the local constraints and expressing some of the fields in terms of the others, while still keeping the locality of the theory. In fact, this is an important part of the Higgs mechanism \cite{Brout1964,Higgs1964} as well as in its lattice version \cite{Fradkin1979}, where the Goldstone bosons are absorbed by the gauge field, giving it a mass, removing all the local constraints, and eliminating part of the matter fields. The unitary transformation that takes care of it is simply the well-known unitary gauge fixing. In this work we will show that a similar procedure can be devised also to eliminate the fermionic matter in the context of $U(N)$ and $SU(N)$ lattice gauge theories (LGTs) \cite{Wilson,KogutSusskind} with staggered fermions in the Hamiltonian representation \cite{Susskind1977}.

In order to obtain this result, we extend the method of Ref. \cite{Fradkin1979} to LGTs where the matter is represented by hard-core bosons in a staggered configuration. Combining that with our recent work \cite{Zohar2018b}, where we showed that fermionic matter in $SU(N)$ and $U(N)$ LGTs can be transformed into hard-core bosons, allows us to completely eliminate the fermionic matter while keeping locality. Specifically, on the one hand we introduce a unitary transformation that decouples the matter from the gauge fields in the Hamiltonian, while keeping the locality of the theory. Furthermore, it enforces a trivial product state for the matter, and a new local constraint for the gauge fields, which stems from the finiteness of the local Hilbert space of the bosons. Thus, even though the original gauge symmetry is broken by explicitly solving the Gauss laws as in the scalar field case \cite{Fradkin1979}, other local constraints satisfied by the gauge fields alone arise. On the other hand, the unitary transformation introduced in \cite{Zohar2018b} replaces the fermionic degrees of freedom by (hard-core) bosonic ones, while transferring the statistics to the gauge field. As a result, the interaction range is slightly, but locally, extended. As this procedure uses only a finite subgroup of the gauge group, the gauge symmetry is not broken and one obtains Gauss laws in which the fermionic degrees of freedom are replaced by hard-core bosonic ones. Thus, the complete procedure consists of two transformations: firstly, the fermionic matter is converted to hard-core bosonic one and secondly, the latter is eliminated. The resulting theory only contains gauge degrees of freedom, is local, breaks local gauge invariance, but introduces other local constrains.

The work is organized as follows. First, we argue that the Gauss law could be solved for matter fields in arbitrary dimensions - while a complete solution of it for the gauge field is not always possible; then we proceed to a demonstration of the $U(1)$ case, starting with a review of the $U(1)$ unitary gauge for the complex scalar field on the lattice of \cite{Fradkin1979}, extend it to hard-core bosonic matter (that has finite local spaces but no fermionic statistics) and eventually  to staggered fermionic matter fields by combining with the results of \cite{Zohar2018b}. After these demonstrations, we generalize the procedure to $U(N)$ and $SU(N)$ lattice gauge theories,  with any $N>1$, coupled to fundamental staggered fermions.

Throughout this work we assume the summation of repeated matrix and vector indices, unless stated differently.

\section{The Gauss Law}

\subsection{The Classical Case}
Already in the context of classical electrodynamics, the gauge fields and matter are related through the Gauss law, that has nothing to do with quantization. It is given by the equation
\begin{equation}
\nabla \cdot \mathbf{E}\left(\mathbf{x}\right) = \rho\left(\mathbf{x}\right)
\label{classical_gauss}
\end{equation}
stating that the divergence of the electric field $\mathbf{E}\left(\mathbf{x}\right)$ at any space point $\mathbf{x}$ is equal to the local charge density $\rho\left(\mathbf{x}\right)$ associated with the matter. It is one of Maxwell's equations, that is obtained as an equation of motion in the Euler-Lagrange formalism; nevertheless, it is a static equation that includes no time derivatives - a set of local constraints. Indeed, when one uses instead the Hamilton formalism that does not treat time and space on an equal footing, it does not appear as an equation of motion anymore (but rather as a constraint that is added to the Hamiltonian with a Lagrange multiplier after the Legendre transformation).

The question is, then, whether we could solve the constraints and use them for reducing the number of degrees of freedom in our system by completely eliminating either the gauge field or the matter. If it is the gauge field that we wish to eliminate, we should solve (\ref{classical_gauss}) for $\mathbf{E}\left(\mathbf{x}\right)$. In this case it is a differential equation, and it can only be integrated in one space dimension where it becomes a simple first order equation, $\partial E / \partial x = \rho\left(x\right)$, giving rise to a nonlocal solution,
\begin{equation}
E\left(x\right) = \overset{x}{\int} dx' \rho\left(x'\right)
\label{sol1d}
\end{equation}
This is a well known solution that may be generalized to non-Abelian groups and quantum cases as well, but not beyond $1+1d$: in more dimensions the electric fields has more components and there are just not enough constraints (constants of motion) that can be used for integration (the electric field is a rotation-vector, and the equations are rotation-scalars).

The solution for the matter, on the other hand, is completely different, as the charge density is a rotation-scalar, and (\ref{classical_gauss}) is just a simple algebraic equation for it, already explicitly and locally solved. In this work we will show how to use that for completely eliminating matter fields in particular lattice gauge theories.

\subsection{Lattice Gauge Theories}
The physical Hilbert space of a quantum gauge theory, $\mathcal{H}_{\text{phys}}$, is contained in the product of the Hilbert spaces of the gauge field, $\mathcal{H}_{\text{gauge}}$, and the matter, $\mathcal{H}_{\text{matter}}$:
\begin{equation}
\mathcal{H}_{\text{phys}} \subset \mathcal{H}_{\text{gauge}} \otimes \mathcal{H}_{\text{matter}}
\end{equation}
It is not equal to the product, as the gauge field and matter degrees of freedom are connected through the Gauss law, that becomes (for electrodynamics) the eigenvalue equation
\begin{equation}
\nabla \cdot \mathbf{E}\left(\mathbf{x}\right)\left|\psi\right\rangle = \rho\left(\mathbf{x}\right)\left|\psi\right\rangle
\end{equation}

From now on, we will focus on lattice gauge theories \cite{Wilson,KogutSusskind}. In the Hamiltonian picture \cite{KogutSusskind} time is continuous, and the degrees of freedom reside, in $d+1$ dimensions, on a $\mathbb{Z}^d$ lattice.  We will begin our demonstration with $U(1)$ gauge fields.

The matter, either bosonic or fermionic, resides on the vertices. Later on we will focus on particular types of matter, but now we it will be enough for us to define charge operators $Q\left(\mathbf{x}\right)$ at each vertex $\mathbf{x} \in \mathbb{Z}^d$. In the $U(1)$ case it will take integer eigenvalues, either bounded or not, depending on the type of matter used. Gauge fields, on the other hand, reside on the lattice's links $\left(\mathbf{x},i\right)$, emanating from the vertex $\mathbf{x}$ in the direction $i=1,...,d$. The local Hilbert space of a $U(1)$ gauge field on a link is this of a particle on a ring: the role of the vector potential will be played by a compact variable $\phi\left(\mathbf{x}\right)$, canonically conjugate to the electric field operator - the "angular momentum" $E\left(\mathbf{x},i\right)$, with an unbounded integer spectrum. Thanks to the canonical relation
\begin{equation}
\left[\phi\left(\mathbf{x},i\right),E\left(\mathbf{y},j\right)\right] = i \delta_{ij} \delta \left(\mathbf{x},\mathbf{y}\right)
\label{phiE}
\end{equation}
(where both deltas are Kronecker's)
the \emph{group element operator} defined by
\begin{equation}
U\left(\mathbf{x}\right) = e^{i \phi\left(\mathbf{x},i\right)}
\end{equation}
is simply an electric field raising operator,
\begin{equation}
\left[E\left(\mathbf{x},i\right),U\left(\mathbf{y},j\right)\right] = \delta_{ij}\delta\left(\mathbf{x},\mathbf{y}\right)U\left(\mathbf{y},j\right)
\end{equation}

The pure-gauge part of the Hamiltonian of such theories is usually the \emph{Kogut-Susskind Hamiltonian},
\begin{widetext}
\begin{equation}
H_{KS} = \frac{g^2}{2}\underset{\mathbf{x},i}{\sum}E^2\left(\mathbf{x},i\right)
-\frac{1}{g^2}\underset{\mathbf{x},i<j}{\sum}\cos\left(\phi\left(\mathbf{x},i\right)+\phi\left(\mathbf{x}+\mathbf{e}_i,j\right)-\phi\left(\mathbf{x}+\mathbf{e}_j,i\right)-\phi\left(\mathbf{x},j\right)\right)
\end{equation}
\end{widetext}
where $\mathbf{e}_i$ is a unit vector in the $i$ direction.

The interaction of the matter with the gauge field takes the form
\begin{equation}
H_{\text{int}} = \underset{\mathbf{x},i}{\sum}\left(\epsilon\left(\mathbf{x},i\right)a^{\dagger}\left(\mathbf{x}\right)U\left(\mathbf{x},i\right)a\left(\mathbf{x}+\mathbf{e}_i\right) + h.c\right)
\end{equation}
where $a^{\dagger}\left(\mathbf{x}\right)$ is bosonic or fermionic, such that
\begin{equation}
\left[Q\left(\mathbf{x}\right),a^{\dagger}\left(\mathbf{y}\right)\right]=\delta\left(\mathbf{x},\mathbf{y}\right)a^{\dagger}\left(\mathbf{x}\right)
\end{equation}
This can be added to some free matter part, $H_M$, and altogether $H=H_{KS}+H_M+H_{\text{int}}$.

Gauge invariance is the invariance under transformations generated by the local generators
\begin{equation}
\mathcal{G}\left(\mathbf{x}\right) = \underset{i}{\sum}\left(E\left(\mathbf{x},i\right) - E\left(\mathbf{x}-\mathbf{e}_i,i\right)\right) - Q\left(\mathbf{x}\right)
\end{equation}
These operators commute with the Hamiltonian,
\begin{equation}
\left[H,\mathcal{G}\left(\mathbf{x}\right)\right]=0 \quad\quad\quad\forall \mathbf{x}\in\mathbb{Z}^d
\label{supersel}
\end{equation}
- a local symmetry, or a set of local constraints. A physical state $\left|\psi\right\rangle$ is gauge invariant, that is, it is an eigenstate of all the local generators $\mathcal{G}\left(\mathbf{x}\right)$,
\begin{equation}
\mathcal{G}\left(\mathbf{x}\right)\left|\psi\right\rangle = \lambda\left(\mathbf{x}\right)\left|\psi\right\rangle
\label{eigval}
\end{equation}
and the commutation relations (\ref{supersel}) imply that states with different $\left\{\lambda\left(\mathbf{x}\right)\right\}$ are not connected by the dynamics, and give rise to disconnected sectors,
\begin{equation}
\mathcal{H}_{\text{phys}} = \bigcup \mathcal{H}_{\text{phys}}\left(\left\{\lambda\left(\mathbf{x}\right)\right\}\right)
\end{equation}

The eigenvalue equation (\ref{eigval}) could be rewritten as
\begin{equation}
\underset{i}{\sum}\left(E\left(\mathbf{x},i\right) - E\left(\mathbf{x}-\mathbf{e}_i,i\right)\right)\left|\psi\right\rangle = \left(Q\left(\mathbf{x}\right)+\lambda\left(\mathbf{x}\right)\right)\left|\psi\right\rangle
\label{Gausslaw}
\end{equation}
- which we recognize as the Gauss law: the divergence of electric fields at a vertex equals the charge there, which is composed of the dynamical charge - the operator $Q\left(\mathbf{x}\right)$, and the eigenvalues $\lambda\left(\mathbf{x}\right)$ which we can now recognize as static charges. From now on we shall choose $\lambda\left(\mathbf{x}\right) = 0$ everywhere - that is, restrict our physical Hilbert space to the sector with no static charges, but the results may be generalized in a straightforward way also to any other charge  sector. Defining the local electric field divergence operator as
$D\left(\mathbf{x}\right) = \underset{i}{\sum}\left(E\left(\mathbf{x},i\right) - E\left(\mathbf{x}-\mathbf{e}_i,i\right)\right)$,
we can now rewrite the Gauss law we will use from now on as
\begin{equation}
D\left(\mathbf{x}\right)\left|\psi\right\rangle = Q\left(\mathbf{x}\right)\left|\psi\right\rangle
\label{GaussD}
\end{equation}

\subsection{Gauge Invariant States}
A general state in $\mathcal{H}_{\text{gauge}} \otimes \mathcal{H}_{\text{matter}}$ could be expanded as
$\underset{g,m}{\sum}A\left(g,m\right)\left|g\right\rangle_{\text{gauge}}\otimes\left|m\right\rangle_{\text{matter}}$, using some arbitrary bases
$\left|g\right\rangle,\left|m\right\rangle$ of the gauge field and matter Hilbert spaces respectively. However, thanks to the Gauss law (\ref{GaussD}), a physical state may be expanded in a more restrictive way. For that, we define the charge states of the matter - eigenstates (not necessarily unique) of the charge operators $\left\{Q\left(\mathbf{x}\right)\right\}$,
\begin{equation}
Q\left(\mathbf{x}\right)\left|\left\{q\right\}\right\rangle = q\left(\mathbf{x}\right)\left|\left\{q\right\}\right\rangle
\end{equation}
Similarly, we can define (non-unique) eigenstates of the electric field divergence operators $D\left(\mathbf{x}\right)$ (not unique) as
\begin{equation}
D\left(\mathbf{x}\right)\left|\left\{d\right\}\right\rangle = d\left(\mathbf{x}\right)\left|\left\{d\right\}\right\rangle
\end{equation}
The gauge invariant states may be expanded as
\begin{equation}
\left|\psi\right\rangle = \underset{\left\{q,d\right\}}{\sum} f\left(\left\{q,d\right\}\right)
\left[\underset{\mathbf{x}}{\prod}\delta_{d\left(\mathbf{x}\right),q\left(\mathbf{x}\right)}\right]\left|\left\{d\right\}\right\rangle_{\text{gauge}}\otimes\left|\left\{q\right\}\right\rangle_{\text{matter}}
\label{psig}
\end{equation}
The notations $\left|\left\{q\right\}\right\rangle$ and $\left|\left\{d\right\}\right\rangle$ are, as mentioned above, generally non-unique: different local matter configurations, corresponding to different quantum states, may give rise to similar local $\left\{q\right\}$ eigenvalues of the charge operators $\left\{Q\right\}$, and similarly with the divergences of the electric field $\left\{D\right\}$ and their eigenvalues $\left\{d\right\}$ (which is the usual case, in which more indices can be added); the notations above are merely illustrative and their accurate details will be later discussed, here the important thing we wanted to emphasize is that the physical Hilbert space is not spanned by all the product states).

\subsection{Solving the Gauss Laws in a Lattice Gauge Theory}

After having described the Hilbert space of a lattice gauge theory, and the implication of the local symmetries on its structure, we are ready to see how to use the Gauss laws, or their explicit solutions, for simplifying the description and reduction of redundant degrees of freedom.

We wish to discuss particular types of unitary transformations, that take a gauge invariant state $\left|\psi\right\rangle$ as in (\ref{psig}), satisfying (\ref{GaussD}), and completely eliminate either the gauge field or matter degrees of freedom, while conserving the  physical information - the amplitudes of elements in superposition. This will be done by using the gauge symmetry and  will, eventually, break it. One can consider, obviously, other types of transformations that leave some ingredients of the field and do not remove them completely, but we will not discuss such transformations here.

A complete removal of the gauge field would be done by a transformation $\mathcal{W}$, as follows:
\begin{equation}
\mathcal{W}\left|\psi\right\rangle = \left|0\right\rangle_{\text{gauge}}\otimes\underset{\left\{q\right\}}{\sum} f\left(\left\{q\right\}\right)
\left|\left\{q\right\}\right\rangle_{\text{matter}}
\label{Wtrans}
\end{equation}
where $\left|0\right\rangle_{\text{gauge}}$ is some "empty" gauge field state that is factored out.

This transformation, in the cases it can be defined, is a controlled operation, making use of the symmetry and the local set of constraints. One locally transforms the degrees of freedom that are to be decoupled, controlled by those that remain. This requires the solution of the Gauss law that was previously discussed; for that reason, the transformation $\mathcal{W}$ is only possible, in general, for one spatial dimension, where the gauge field can be integrated, using the quantum, lattice analogue of (\ref{sol1d}). Just like there the divergence of a vector quantity was replaced by a simple differential equation, here we obtain a simple difference equation, that can be inverted non-locally, since the number of links (electric fields) is equal to this of vertices (charges and constraints), while in more dimension there are not enough equations: in $d+1$ dimensions, the links scale as $d$ times the vertices. Therefore, Only in this case the gauge field is completely redundant - in more dimensions it cannot be completely removed. In $1+1d$ we define (for open boundary conditions - periodic ones do not allow to remove the gauge field completely)
\begin{equation}
\mathcal{W}=\exp\left(i\underset{x}{\sum}\phi\left(x\right)\underset{y<x}{\sum}Q\left(y\right)\right)
\end{equation}
Using the canonical relation (\ref{phiE}), one obtains that
\begin{equation}
\mathcal{W}E\left(x\right)\mathcal{W}^{\dagger} = E\left(x\right) - \underset{y<x}{\sum}Q\left(y\right)
\end{equation}
which is zero on the physical Hilbert space, thanks to the Gauss laws. This type of transformation was used, for example, in \cite{Hamer1997,Bringoltz2009,Banuls2013,Martinez2016,Sala2018}. For a further discussion and the relation to minimal coupling, refer to \cite{Zohar2018c}.

In the next sections we will focus on the other option, where the matter is decoupled and eliminated.

\section{Eliminating the Matter from a $U(1)$ Lattice Gauge Theory}

We wish to construct a unitary transformation $\mathcal{U}$ which similarly decouples the matter from a physical state $\left|\psi\right\rangle$
as in (\ref{psig}), satisfying the Gauss law (\ref{GaussD}) - that is, a transformation of the type
\begin{equation}
\mathcal{U}\left|\psi\right\rangle = \left|0\right\rangle_{\text{matter}}\otimes\underset{\left\{d\right\}}{\sum} f\left(\left\{d\right\}\right)
\left|\left\{d\right\}\right\rangle_{\text{gauge}}
\label{Utrans}
\end{equation}
Below we shall see when such a transformation can be constructed and how. It will involve the solution of the Gauss law for the matter, and therefore, as discussed above, it does not depend on the dimension of space: the number of equations is exactly the number of charges, or the number of electric field divergences.

Let us assume the existence of unitary transformations $u\left(\mathbf{x}\right)$ that act on the matter at the vertex $\mathbf{x}$. They mutually commute,
\begin{equation}
\left[u\left(\mathbf{x}\right),u\left(\mathbf{y}\right)\right]=\left[u\left(\mathbf{x}\right),u^{\dagger}\left(\mathbf{y}\right)\right]=0
\label{ucomm}
\end{equation} and change the local charges as follows:
\begin{equation}
u^{q}\left|q\right\rangle = \left|0\right\rangle
\label{udef0}
\end{equation}
for each integer $q \neq 0$. Since $u$ is unitary, negative values involve hermitian conjugation.

We still have to check in which cases such transformations exist; but when they do, in order to eliminate the charge, we will have to act on each vertex with a power of this transformation, that is exactly equal to the initial amount of charge that was there. We can do it thanks to the Gauss law; let us define a controlled, local unitary transformation $\mathcal{U}\left(\mathbf{x}\right)$ that lowers the charge at $\mathbf{x}$ by an amount given by $D\left(\mathbf{x}\right)$, the divergence of electric field there. Thanks to the Gauss law (\ref{GaussD}), this is equal to the charge and hence, acting on gauge invariant states it will reduce $Q\left(\mathbf{x}\right)$ to zero. Let us see that explicitly; consider an expansion (\ref{psig}) of a gauge invariant state $\left|\psi\right\rangle$. Then,
\begin{equation}
\begin{aligned}
&Q\left(\mathbf{x}\right)u\left(\mathbf{x}\right)^{D\left(\mathbf{x}\right)}\left|\psi\right\rangle = \\
& \underset{\left\{q,d\right\}}{\sum} f\left(\left\{q\right\}\right)
\left[\underset{\mathbf{x}}{\prod}\delta_{d\left(\mathbf{x}\right),q\left(\mathbf{x}\right)}\right]u\left(\mathbf{x}\right)^{D\left(\mathbf{x}\right)}\left|\left\{d\right\}\right\rangle_{\text{gauge}}\otimes\left|\left\{q\right\}\right\rangle_{\text{matter}}=\\
& \underset{\left\{q,d\right\}}{\sum} f\left(\left\{q\right\}\right)
\left[\underset{\mathbf{x}}{\prod}\delta_{d\left(\mathbf{x}\right),q\left(\mathbf{x}\right)}\right]u\left(\mathbf{x}\right)^{d\left(\mathbf{x}\right)}\left|\left\{d\right\}\right\rangle_{\text{gauge}}\otimes\left|\left\{q\right\}\right\rangle_{\text{matter}}=\\
& \underset{\left\{q,d\right\}}{\sum} f\left(\left\{q\right\}\right)
\left[\underset{\mathbf{x}}{\prod}\delta_{d\left(\mathbf{x}\right),q\left(\mathbf{x}\right)}\right]\left|\left\{d\right\}\right\rangle_{\text{gauge}}\otimes
u\left(\mathbf{x}\right)^{q\left(\mathbf{x}\right)}
\left|\left\{q\right\}\right\rangle_{\text{matter}}\\
&=0
\end{aligned}
\end{equation}

We can hence define the local controlled unitaries
\begin{equation}
\mathcal{U}\left(\mathbf{x}\right) = u\left(\mathbf{x}\right)^{D\left(\mathbf{x}\right)}
\label{Uxdef}
\end{equation}
and the transformation
\begin{equation}
\mathcal{U}=\underset{\mathbf{x}}{\prod}\mathcal{U}\left(\mathbf{x}\right)
\label{Udef}
\end{equation}
will give rise to (\ref{Utrans}), as we wish.

All this can be achieved, for example, if $u$ is a unitary lowering charge operator,
\begin{equation}
\left[Q\left(\mathbf{x}\right),u\left(\mathbf{x}\right)\right]=-u\left(\mathbf{x}\right)
\label{udef}
\end{equation}
but it is not required. In such a case, an infinite charge ladder is required, as in the case of complex scalar fields which will be the first we discuss.
How does the Gauss law transform in such a case? It is simple to see in the case that $u$ is a charge lowering operator, e.g. that (\ref{udef}) is satisfied. Then, since $\mathcal{U} D\left(\mathbf{x}\right) \mathcal{U}^{\dagger}=D\left(\mathbf{x}\right)$ , and
\begin{equation}
\mathcal{U} Q\left(\mathbf{x}\right) \mathcal{U}^{\dagger} = Q\left(\mathbf{x}\right)+D\left(\mathbf{x}\right)
\label{Qtrans}
\end{equation}
we obtain that $\mathcal{U} \left(D\left(\mathbf{x}\right)-Q\left(\mathbf{x}\right)\right) \mathcal{U}^{\dagger} = 0$, and hence the Gauss law (\ref{GaussD}) transforms to a trivial $0=0$ and the gauge symmetry completely breaks down.
When (\ref{udef}) is not satisfied, one has to be extra cautious, and this, as we shall see, will be the case when the matter is fermionic.

Next, let us see in which cases such transformations could be defined.

\subsection{Complex Scalar Matter}

First, we consider the case in which each vertex hosts a complex scalar field, $\Phi\left(\mathbf{x}\right)$, for which the relevant transformation - the lattice version of the unitary gauge of the Brout-Englert-Higgs mechanism \cite{Brout1964,Higgs1964} was discussed by Fradkin and Shenker in \cite{Fradkin1979}. The field may be expanded in a polar form, with a radial and angular part (phase); we assume, for simplicity, that the radial part is frozen, for example due to the Higgs mechanism.  Following \cite{Fradkin1979},
\begin{equation}
\Phi\left(\mathbf{x}\right) = R_0 e^{-i \theta\left(\mathbf{x}\right)}
\end{equation}
At each vertex we have a non-bounded charge operator $Q\left(\mathbf{x}\right)$ with an infinite spectrum of integers. It is raised by the operators $e^{i\theta\left(\mathbf{x}\right)}$, which mutually commute, and therefore
\begin{equation}
u\left(\mathbf{x}\right)=e^{-i\theta\left(\mathbf{x}\right)}
\end{equation}
satisfies both (\ref{udef}) and (\ref{ucomm}), allowing us to obtain a well defined $\mathcal{U}$ using (\ref{Uxdef}) and (\ref{Udef}). Since $u$ here is a unitary lowering operator, the Gauss laws will vanish for the transformed state using (\ref{Qtrans}).

The complete transformation then takes the form
\begin{equation}
\mathcal{U}=e^{-i\underset{\mathbf{x}}{\sum}\theta\left(\mathbf{x}\right)D\left(\mathbf{x}\right)}
=e^{i\underset{\mathbf{x},i}{\sum}E\left(\mathbf{x},i\right)\left(\theta\left(\mathbf{x}+\mathbf{e}_i\right)-\theta\left(\mathbf{x}\right)\right)}
\end{equation}
which, seen now as a transformation of the gauge fields controlled by the matter, is recognized as the unitary gauge \cite{Fradkin1979}. In such theories, the interaction terms take the form $\Phi^{\dagger}\left(\mathbf{x}\right)U\left(\mathbf{x},i\right)\Phi\left(\mathbf{x}+\mathbf{e}_i\right)$, which, after freezing the radial field, become (if $\epsilon$ is real)
\begin{equation}
H_{\text{int}} = 2R_0^2\underset{\mathbf{x},i}{\sum}\epsilon\left(\mathbf{x},i\right)\cos\left(\phi\left(\mathbf{x},i\right) + \theta\left(\mathbf{x}\right) - \theta\left(\mathbf{x}+\mathbf{e}_i\right)\right)
\end{equation}
Note that
\begin{equation}
\mathcal{U}\phi\left(\mathbf{x},i\right)\mathcal{U}^{\dagger} = \phi\left(\mathbf{x},i\right) +\theta\left(\mathbf{x}+\mathbf{e}_i\right) - \theta\left(\mathbf{x}+\mathbf{e}_i\right)
\end{equation}
- therefore, after the transformation the matter field decouples from the interaction terms, and we obtain massive gauge fields:
\begin{equation}
\tilde{H}_{\text{int}} = \mathcal{U}H_{\text{int}}\mathcal{U}^{\dagger} = 2R_0^2\underset{\mathbf{x},i}{\sum}\epsilon\left(\mathbf{x},i\right)\cos\left(\phi\left(\mathbf{x},i\right)\right)
\end{equation}
 thanks to the terms proportional to
$R_0^2 \cos\left(\phi\left(\mathbf{x},i\right)\right)$. These terms explicitly break gauge invariance, that does not exist anymore as anticipated. The other parts of the Hamiltonian commute with the transformation and do not transform (the rest are pure gauge terms, and the $G\left(\mathbf{x}\right)$ are generators of pure gauge transformations and therefore commute with them).

Similar transformations are possible for other Higgs scenarios, with different groups. For example, see \cite{Haegeman2014} for $\mathbb{Z}_2$. If the radial component of the field is not frozen, it will not be eliminated, since it is not coupled to the gauge field and thus not subject to any local constraints.

Before we move on to other types of matter, one could ask what happens if we couple the same gauge field to more matter fields, residing at the vertices and adding up to the local charges, $Q\left(\mathbf{x}\right) = \underset{i}{\sum}Q_i\left(\mathbf{x}\right)$. In this case, in general, a transformation of the form $\mathcal{U}$ will not be possible, since the spectrum of the local charge operators $Q\left(\mathbf{x}\right)$ will be degenerate. Microscopical matter configurations must be completely distinguishable in terms of their charges, if one wishes to decouple the matter in the manner described above. The controlled operation is based on the divergence of the electric fields which is equal to the total fermionic charge at the vertex, with no way to distinguish different charge contributions.

\subsection{Staggered Hard-Core Bosonic matter}
Our next stop en route to fermions will be hard-core bosonic matter. In this case, the matter Hilbert space on each vertex is that of a spin-half particle. To be able to eventually generalize to  staggered fermions, we will stagger the hard-core bosons. On each vertex we define the "number operator",
\begin{equation}
n\left(\mathbf{x}\right) = \frac{1}{2}\left(\sigma_z\left(\mathbf{x}\right)+1\right)
\end{equation}
and the staggered charge operators
\begin{equation}
Q\left(\mathbf{x}\right) = n\left(\mathbf{x}\right) - s\left(\mathbf{x}\right)
\label{Qdef}
\end{equation}
where $s\left(\mathbf{x}\right)=0$ ($1$) on the even (odd) sublattice representing particles (anti-particles). The Gauss law (\ref{GaussD}) may then be rewritten as
\begin{equation}
G\left(\mathbf{x}\right)\left|\psi\right\rangle = n\left(\mathbf{x}\right)\left|\psi\right\rangle
\label{GaussG}
\end{equation}
introducing
\begin{equation}
G\left(\mathbf{x}\right)=D\left(\mathbf{x}\right) + s\left(\mathbf{x}\right)
\label{Gdef}
\end{equation}

The parts of the Hamiltonian that depend on the matter take the form
\begin{equation}
H_M = M\underset{\mathbf{x}}{\sum}\left(-1\right)^{s\left(\mathbf{x}\right)}n\left(\mathbf{x}\right)
\end{equation}
and
\begin{equation}
H_{\text{int}} = \underset{\mathbf{x},i}{\sum}\left(\epsilon\left(\mathbf{x},i\right)\sigma_{+}\left(\mathbf{x}\right)U\left(\mathbf{x},i\right)\sigma_{-}\left(\mathbf{x}+\mathbf{e}_i\right) + h.c\right)
\end{equation}

Unlike in the scalar case, now the charges are bounded, because the operators $n\left(\mathbf{x}\right)$ are bounded. The Gauss law in the form (\ref{GaussG}) implies that the spectrum of the operators $G\left(\mathbf{x}\right)$ in the physical Hilbert space consists only of $0,1$. Therefore we can write down extra local constraints that, at this point, are completely redundant:
\begin{equation}
G\left(\mathbf{x}\right)\left(G\left(\mathbf{x}\right)-1\right)\left|\psi\right\rangle = 0
\label{GG}
\end{equation}
We define, on each vertex $\mathbf{x}$, the operators $P_{g}\left(\mathbf{x}\right)$, which project to the subspaces where $G\left(\mathbf{x}\right)=g$. The physical Hilbert space is contained within the subspace of $g=0,1$ everywhere, and therefore we can multiply the entire Hamiltonian by the projectors $P\left(\mathbf{x}\right) = P_1\left(\mathbf{x}\right)+P_0\left(\mathbf{x}\right)$ and have exactly the same spectrum and dynamics \emph{within the physical Hilbert space}. However, since most of the Hamiltonian terms commute with the operators $G\left(\mathbf{x}\right)$, it is sufficient to do it for the interaction part - the only part that does not commute - and even there it is sufficient to include only the most relevant local projectors. Finally we obtain, using the Gauss law, that within the physical Hilbert space,
\begin{widetext}
\begin{equation}
H^{\text{phys}}_{\text{int}} = \underset{\mathbf{x},i}{\sum}\left(\epsilon\left(\mathbf{x},i\right)P_{1}\left(\mathbf{x}\right)\sigma_{+}\left(\mathbf{x}\right)U\left(\mathbf{x},i\right)\sigma_{-}\left(\mathbf{x}+\mathbf{e}_i\right) P_{0}\left(\mathbf{x}+\mathbf{e}_i\right) + h.c\right)
\label{Hspinphys}
\end{equation}

The Gauss law also helps us to rewrite, within the physical Hilbert space, the mass part of the Hamiltonian as
\begin{equation}
H^{\text{phys}}_M = M\underset{\mathbf{x}}{\sum}\left(-1\right)^{s\left(\mathbf{x}\right)}G\left(\mathbf{x}\right) = 2M\underset{\mathbf{x},i}{\sum}\left(-1\right)^{s\left(\mathbf{x}\right)}
E\left(\mathbf{x},i\right) + \text{const.}
\label{HMM}
\end{equation}
\end{widetext}
Another implication is that now, that we do not have an infinite ladder of charges, we cannot define a unitary $u$ that lowers $Q$. Instead, we can define
\begin{equation}
u\left(\mathbf{x}\right) = \sigma_x\left(\mathbf{x}\right)
\end{equation}
which will satisfy (\ref{udef0}) but not (\ref{udef}).

We would like to construct a local transformation $\mathcal{U}\left(\mathbf{x}\right)$ that eliminates the charges. Using the modified, staggered Gauss law (\ref{GaussG}), we construct the local transformation
\begin{equation}
\mathcal{U}\left(\mathbf{x}\right) = u\left(\mathbf{x}\right)^{G\left(\mathbf{x}\right)}=\sigma_x\left(\mathbf{x}\right)^{G\left(\mathbf{x}\right)}
\end{equation}
That takes a spin up state (corresponding thanks to the Gauss law to $G=1$) to spin down, and leaves spin down invariant. The operators $G\left(\mathbf{x}\right)$ are left invariant, and using $\mathcal{U}\left(\mathbf{x}\right) \sigma_z\left(\mathbf{x}\right) \mathcal{U}^{\dagger}\left(\mathbf{x}\right) = \left(-1\right)^{G\left(\mathbf{x}\right)} \sigma_z\left(\mathbf{x}\right)$ we obtain that the transformed state, $\left|\tilde \psi\right\rangle = \mathcal{U}\left|\psi\right\rangle$ satisfies
\begin{equation}
\sigma_z\left(\mathbf{x}\right) \left|\tilde \psi\right\rangle = - \left(-1\right)^{G\left(\mathbf{x}\right)} \left(1-2G\left(\mathbf{x}\right)\right)\left|\tilde\psi\right\rangle
\label{ztrans}
\end{equation}
From this equation, it looks as if there is still some local coupling between the gauge field and the matter. However, this is not the case; recall the conditions (\ref{GG}) on the spectrum of the $G\left(\mathbf{x}\right)$ operators that were redundant before the transformation; now they are not redundant anymore, and in fact, they are invariant under the transformation, that is
\begin{equation}
G\left(\mathbf{x}\right)\left(G\left(\mathbf{x}\right)-1\right)\left|\tilde\psi\right\rangle = 0
\label{GGG}
\end{equation}
This implies that
\begin{equation}
\left(1-2G\left(\mathbf{x}\right)\right)\left|\tilde\psi\right\rangle = \left(-1\right)^{G\left(\mathbf{x}\right)}\left|\tilde\psi\right\rangle
\end{equation}
and (\ref{ztrans}) simplifies to
\begin{equation}
\sigma_z\left(\mathbf{x}\right) \left|\tilde \psi\right\rangle = -\left|\tilde \psi\right\rangle
\end{equation}
as expected, implying that all the matter degrees of freedom are decoupled and in a spin-down state, or that
\begin{equation}
n\left(\mathbf{x}\right) \left|\tilde \psi\right\rangle = 0
\end{equation}

How does the Hamiltonian transform? As before, the pure gauge part $H_{KS}$ does not transform at all, as it commutes with the transformation. The mass part (\ref{HMM}) commutes as well and does not transform, and we are left with the transformation of the interaction part.
However, in this case we have to be more careful with the transformation of the interaction part of the Hamiltonian that in the complex scalar case. The interaction term does not commute with $G\left(\mathbf{x}\right)$ and thus changes their eigenvalues. Unlike in the scalar case, here we started with matter that resides in finite local Hilbert spaces, giving rise to the local constraints (\ref{GG},\ref{GGG}), which have to be incorporated into the interaction explicitly before transforming it, otherwise we will get terms that breaks them. Thus, we transform the \emph{physical} interaction Hamiltonian, within the physical subspace (\ref{Hspinphys}), and obtain
\begin{widetext}
\begin{equation}
\tilde{H}^{\text{phys}}_{\text{int}} = \mathcal{U}H^{\text{phys}}_{\text{int}}\mathcal{U}^{\dagger} =
\underset{\mathbf{x},i}{\sum}\left(\epsilon\left(\mathbf{x},i\right)P_{1}\left(\mathbf{x}\right)U\left(\mathbf{x},i\right) P_{0}\left(\mathbf{x}+\mathbf{e}_i\right)\otimes \left|\downarrow\right\rangle\left\langle\downarrow\right|_{\mathbf{x}} \otimes \left|\downarrow\right\rangle\left\langle\downarrow\right|_{\mathbf{x}+\mathbf{e}_i} + h.c\right)
\label{spinintrans}
\end{equation}

We can now completely forget about the matter degrees of freedom as they are decoupled. The final Hamiltonian to be used in the transformed physical Hilbert space is, therefore,
\begin{equation}
\underset{\mathbf{x}}{\otimes}\left\langle\downarrow\right|_{\mathbf{x}}
\mathcal{U}H^{\text{phys}}\mathcal{U}^{\dagger}
\underset{\mathbf{x}}{\otimes}\left|\downarrow\right\rangle_{\mathbf{x}}
=
H_{KS} + 2M\underset{\mathbf{x},i}{\sum}\left(-1\right)^{s\left(\mathbf{x}\right)}
E\left(\mathbf{x},i\right)
+
\underset{\mathbf{x},i}{\sum}\left(\epsilon\left(\mathbf{x},i\right)P_{1}\left(\mathbf{x}\right)U\left(\mathbf{x},i\right) P_{0}\left(\mathbf{x}+\mathbf{e}_i\right)+ h.c\right)
\end{equation}
\end{widetext}
without any local Gauss laws but with the local constraints (\ref{GGG}) that are taken care of by the projectors. These projectors extend the range of interaction, and this is the price we have to pay for having, originally, local matter spaces that are finite.

Here, once again, we could not add multiple matter species coupled to the same gauge field, because this would destroy the unique mapping between a charge configuration and a matter state. This also gives a good motivation for staggering: if, instead, we had two hard-core bosonic species per site, coupled to the same gauge field, we would not be able to perform such a decoupling transformation.

\subsection{Staggered Fermionic Matter}
Finally, we are ready to deal with staggered fermionic matter \cite{Susskind1977}. In this case, at each vertex there is one fermionic species, created by the operator $\psi^{\dagger}\left(\mathbf{x}\right)$. The fermionic number operators, as usual, are $n\left(\mathbf{x}\right) = \psi^{\dagger}\left(\mathbf{x}\right)\psi\left(\mathbf{x}\right)$, and with them one may define the charges just like in the hard-core bosonic case (\ref{Qdef}).

The parts of the Hamiltonian that involve the matter take the form
\begin{equation}
H_M = M\underset{\mathbf{x}}{\sum}\left(-1\right)^{s\left(\mathbf{x}\right)}n\left(\mathbf{x}\right)
\end{equation}
and
\begin{equation}
H_{\text{int}} = \underset{\mathbf{x},i}{\sum}\left(\epsilon\left(\mathbf{x},i\right)\psi^{\dagger}\left(\mathbf{x}\right)U\left(\mathbf{x},i\right)\psi\left(\mathbf{x}+\mathbf{e}_i\right) + h.c\right)
\end{equation}

One could be tempted to use a Majorana mode as the local unitary, $u\left(\mathbf{x}\right)=\psi\left(\mathbf{x}\right) + \psi^{\dagger}\left(\mathbf{x}\right)$, and indeed it is unitary and satisfies (\ref{udef0}), but as it has an odd fermionic parity (it changes the parity of the states it acts upon), the commutation property (\ref{ucomm}) does not hold and one cannot define (\ref{Udef}) with it, since the local terms will not have a well defined fermionic parity and hence their product will have to be in some fixed order, giving rise to nonlocal strings.

In a recent work \cite{Zohar2018b} we have shown that lattice gauge theories with staggered fermionic matter whose gauge group contains $\mathbb{Z}_2$ as a normal subgroup may be mapped to lattice gauge theories with hard-core bosonic matter (spins), using a local and unitary transformation that does not involve nonlocal strings. This was done as well thanks to the fact that in a gauge theory one has local constraints, that allow to transfer the statistics information to the gauge fields. However, since in that case only the parity is discussed, and it has to do with the finite $\mathbb{Z}_2$ group that is not continuous, the procedure carried out there did not break the symmetry and only allowed to replace the fermionic matter by hard-core bosons, with the only "payment" of extra, but local, appearances of signs of electric fields in the Hamiltonian, that account for the fermionic statistics.

Since $U(1)$ contains $\mathbb{Z}_2$ as a normal subgroup, one can replace the fermions in a $U(1)$ gauge theory by hard-core bosons. Following \cite{Zohar2018b}, we obtain
that our model is equivalently described by the Hamiltonian $H' = H'_{KS}+H'_{\text{int}}+H'_{M}$, in which $H'_M=H_M$ (but with the spin $n$ operators),
\begin{widetext}
\begin{equation}
H'_{KS} = \frac{g^2}{2}\underset{\mathbf{x},i}{\sum}E^2\left(\mathbf{x},i\right)
-\frac{1}{g^2}\underset{\mathbf{x},i<j}{\sum}\xi_{\mathfrak{p}}\left(\mathbf{x},i,j\right)\cos\left(\phi\left(\mathbf{x},i\right)+\phi\left(\mathbf{x}+\mathbf{e}_i,j\right)-\phi\left(\mathbf{x}+\mathbf{e}_j,i\right)-\phi\left(\mathbf{x},j\right)\right)
\end{equation}
\end{widetext}
and
\begin{equation}
H_{\text{int}} = \underset{\mathbf{x},i}{\sum}\left(\xi\left(\mathbf{x},i\right)\epsilon\left(\mathbf{x},i\right)\sigma_{+}\left(\mathbf{x}\right)U\left(\mathbf{x},i\right)\sigma_{-}\left(\mathbf{x}+\mathbf{e}_i\right) + h.c\right)
\end{equation}
where $\xi_{\mathfrak{p}}\left(\mathbf{x},i,j\right),\xi\left(\mathbf{x},i\right)$ are local functions of the electric fields on links that belong to the plaquette/link they are associated with and its neighboring links \cite{Zohar2018b}. Since these functions commute with $G\left(\mathbf{x}\right)$, one may use the procedure introduced for staggered hard-core bosonic matter to eliminate the matter in this fermionic case as well, once the fermions are converted to hard-core bosons.

\section{Extension to non-Abelian Cases}

After having stated the procedure that allows one to eliminate the matter of $U(1)$ lattice gauge theories, with either hard-core bosonic or fermionic matter, we will now generalize it to non-Abelian groups. For that, let us briefly review the structure of those theories.

\subsection{$SU(N)$ and $U(N)$ Lattice Gauge Theories}

\subsubsection{The groups $U(N)$ and $SU(N)$ and their Cartan subalgebra}
We consider here lattice gauge theories whose gauge group $G$ is either $U(N)$ or $SU(N)$. We denote group elements by $g \in G$, and irreducible representations by $j$; unitary matrix representations are given by the Wigner matrices, $D^j_{mn}\left(g\right)$. As these are Lie groups, the representation $j$ is generated by a set of matrix generators, $\tau^j_a$, whose dimension is referred to as the representation's dimension $\text{dim}\left(j\right)$. For $SU(N)$, there are $N^2-1$ such generators, satisfying the group's Lie algebra
\begin{equation}
\left[\tau^j_a,\tau^j_b\right]=if_{abc}\tau^j_c
\label{TLie}
\end{equation}
all of which Hermitian and traceless matrices.
We shall discuss the \emph{fundamental representation}, whose dimension for $SU(N)$ and $U(N)$ is $N$, and in this case the representation index will be omitted. In this representation, one usually chooses the \emph{Cartan-Weyl} basis for the generators, and the normalization condition
\begin{equation}
\text{Tr}\left(\tau_a \tau_b\right) = \frac{1}{2}\delta_{ab}
\label{trTT}
\end{equation}
follows.

In order to obtain the algebra of $U(N)$ from that of $SU(N)$, one only has to introduce one extra generator,
\begin{equation}
\tau_0 = \frac{1}{\sqrt{2N}}\mathbf{1}
\end{equation}
The normalization is chosen in accordance with (\ref{trTT}).

$SU(N)$ and $U(N)$ $(N>1)$ are non-Abelian Lie groups, whose generators, in general, do not commute, as in (\ref{TLie}). However, there exists a maximal subset of generators that commute - forming the \emph{Cartan subalgebra}. For $SU(N)$ there are $N-1$ such generators - the maximal number of $N$ dimensional diagonal traceless matrices. We shall denote them by $\left\{T_{\mu}\right\}_{\mu=1}^{N-1}$. Since the identity matrix trivially commutes with any other matrix, the Cartan subalgebra of $U(N)$ will include the $N-1$ elements of that of $SU(N)$, as well as $T_0=\tau_0$ - altogether $N$ mutually commuting generators.

\subsubsection{The gauge field}

Let us begin with the description of the gauge degrees of freedom. As in the Abelian case, they reside on the links of the lattice. The gauge field on each link is described by a set of operators \cite{KogutSusskind,Zohar2015}: the \emph{group element operators} $U^j_{mn}$, matrices of gauge field operators that transform as group elements in the $j$ representation (in the fundamental representation, where we omit $j$, these are $N\times N$ matrices), and the \emph{left and right transformation generators}, $L_a$ and $R_a$ respectively. These are two independent sets of operators that fulfill the group's Lie algebra,
\begin{equation}
\begin{aligned}
&\left[L_a,L_b\right]=-if_{abc}L_c \\
&\left[R_a,R_b\right]=if_{abc}R_c \\
&\left[L_a,R_b\right]=0
\end{aligned}
\end{equation}
that generate transformations of the $U^j_{mn}$:
\begin{equation}
\begin{aligned}
&\left[L_{a},U^{j}_{mn}\right]=\left(\tau^j_{a}\right)_{mm'}U^j_{m'n}\\
&\left[R_{a},U^{j}_{mn}\right]=U^j_{mn'}\left(\tau^j_{a}\right)_{n'n}
\end{aligned}
\label{ULR}
\end{equation}
These operators satisfy $R_aR_a = L_aL_a \equiv \mathbf{J}^2$.
We will consider, from now on, only the fundamental representation for the gauge field.

The pure gauge (Kogut-Susskind) part of the Hamiltonian takes the form \cite{KogutSusskind}
\begin{widetext}
\begin{equation}
H_{KS} = \frac{g^2}{2}\underset{\mathbf{x},i}{\sum}\mathbf{J}^2\left(\mathbf{x},i\right)
-\frac{1}{2g^2}\underset{\mathbf{x},i<j}{\sum}\left(\text{Tr}\left(U\left(\mathbf{x},i\right)U\left(\mathbf{x}+\mathbf{e}_i,j\right)U^{\dagger}\left(\mathbf{x}+\mathbf{e}_j,i\right)U^{\dagger}\left(\mathbf{x},j\right)\right)+h.c.\right)
\end{equation}
\end{widetext}

\subsubsection{The matter}
As usual, the matter resides on the vertices. We will consider once again staggered matter, and will restrict our discussion to the fundamental representation, where on each vertex there is a spinor with $N$ components. In normal cases it is fermionic, with creation operators $\psi^{\dagger}_m\left(\mathbf{x}\right)$ satisfying the fermionic algebra
\begin{equation}
\begin{aligned}
\left\{\psi_m\left(\mathbf{x}\right),\psi^{\dagger}_n\left(\mathbf{x}\right)\right\}&=\delta_{mn}\delta\left(\mathbf{x},\mathbf{y}\right)\\
\left\{\psi_m\left(\mathbf{x}\right),\psi_n\left(\mathbf{x}\right)\right\}&=0
\end{aligned}
\end{equation}

Out of the $N$ fermionic species, one constructs the local charges $Q_a\left(\mathbf{x}\right)$. In the $SU(N)$ case, there are $N^2-1$ such charges,
\begin{equation}
Q_a\left(\mathbf{x}\right)=\psi_m^{\dagger}\left(\mathbf{x}\right)\left(\tau_a\right)_{mn}\psi_n\left(\mathbf{x}\right)\quad\quad\quad a=1,...,N^2-1
\end{equation}
In the $U(N)$ case, one adds the extra $U(1)$ charge which is explicit,
\begin{equation}
\begin{aligned}
Q_0\left(\mathbf{x}\right)&=\psi_m^{\dagger}\left(\mathbf{x}\right)\left(\tau_0\right)_{mn}\psi_n\left(\mathbf{x}\right)-\sqrt{\frac{N}{2}}s\left(\mathbf{x}\right)\\&=
\sqrt{\frac{N}{2}}\left(\underset{m}{\sum}n_m\left(\mathbf{x}\right)/N-s\left(\mathbf{x}\right)\right)
\end{aligned}
\end{equation}
The charges $Q_a$ satisfy, on each vertex, the group algebra.

The parts of the Hamiltonian that involve the matter take the form
\begin{equation}
H_M = M\underset{\mathbf{x},m}{\sum}\left(-1\right)^{s\left(\mathbf{x}\right)}n_m\left(\mathbf{x}\right)
\end{equation}
and
\begin{equation}
H_{\text{int}} = \underset{\mathbf{x},i}{\sum}\left(\epsilon\left(\mathbf{x},i\right)\psi_m^{\dagger}\left(\mathbf{x}\right)U_{mn}\left(\mathbf{x},i\right)\psi_n\left(\mathbf{x}+\mathbf{e}_i\right) + h.c\right)
\end{equation}

\subsubsection{Gauge invariance}
Gauge invariance now is non-Abelian: there are $N^2-1$ local generators of gauge symmetry at each vertex for $SU(N)$, and one more for $U(N)$. As the groups are non-Abelian, these do not commute. One defines the non-Abelian divergence of electric fields by
\begin{equation}
D_a\left(\mathbf{x}\right)=\underset{i}{\sum}\left(L_a\left(\mathbf{x},i\right) - R_a\left(\mathbf{x}+\mathbf{e}_i,i\right)\right)
\end{equation}

The Hamiltonian is invariant under transformations generated by $D_a\left(\mathbf{x}\right)-Q_a\left(\mathbf{x}\right)$; the non-Abelian Gauss law is
\begin{equation}
D_a\left(\mathbf{x}\right)\left|\psi\right\rangle=Q_a\left(\mathbf{x}\right)\left|\psi\right\rangle
\label{GNA}
\end{equation}
- $R_a,L_a$ play the role of non-Abelian, right and left electric fields.

\subsection{Eliminating the matter in the $U(N)$ case}
As in the $U(1)$ case, for the elimination of matter we will first convert it to a hard-core bosonic form, following \cite{Zohar2018b}. The mapping to bosons is possible in the $U(N)$ case, as it includes $\mathbb{Z}_2$ as a normal subgroup. The fermions are mapped to spinors $\eta_m^{\dagger}\left(\mathbf{x}\right)$, whose components anti-commute on-site,
\begin{equation}
\begin{aligned}
\left\{\eta_m\left(\mathbf{x}\right),\eta_n^{\dagger}\left(\mathbf{x}\right)\right\}&=\delta_{mn}\\
\left\{\eta_m\left(\mathbf{x}\right),\eta_n\left(\mathbf{x}\right)\right\}&=0
\end{aligned}
\end{equation}
but commute between different sites,
\begin{equation}
\left[\eta_m\left(\mathbf{x}\right),\eta_n^{\dagger}\left(\mathbf{y}\right)\right]=
	 \left[\eta_m\left(\mathbf{x}\right),\eta_n\left(\mathbf{y}\right)\right]=0
\end{equation}
for $\mathbf{x}\neq\mathbf{y}$.

Number operators become
\begin{equation}
n_{m}\left(\mathbf{x}\right) = \eta^{\dagger}_m\left(\mathbf{x}\right)\eta_m\left(\mathbf{x}\right)
\end{equation}
and the charges are
\begin{equation}
\begin{aligned}
Q_a\left(\mathbf{x}\right)&=\eta_m^{\dagger}\left(\mathbf{x}\right)\left(T_a\right)_{mn}\eta_n\left(\mathbf{x}\right)\quad\quad\quad a=1,...,N^2-1\\
Q_0\left(\mathbf{x}\right)&=
\sqrt{\frac{N}{2}}\left(\underset{m}{\sum}n_m\left(\mathbf{x}\right)/N-s\left(\mathbf{x}\right)\right)
\end{aligned}
\end{equation}

The Hamiltonian transforms to $H'=H'_{KS}+H'_{\text{int}}+H'_M$, where $H'_M=H_M$ (but with the new definition of number operators with the hard-core bosons),
\begin{widetext}
	\begin{equation}
	H'_{KS} = \frac{g^2}{2}\underset{\mathbf{x},i}{\sum}\mathbf{J}^2\left(\mathbf{x},i\right)
	 -\frac{1}{2g^2}\underset{\mathbf{x},i<j}{\sum}\left(\xi_{\mathfrak{p}}\left(\mathbf{x},i,j\right)\text{Tr}\left(U\left(\mathbf{x},i\right)U\left(\mathbf{x}+\mathbf{e}_i,j\right)U^{\dagger}\left(\mathbf{x}+\mathbf{e}_j,i\right)U^{\dagger}\left(\mathbf{x},j\right)\right)+h.c.\right)
	\end{equation}
\end{widetext}
and
\begin{equation}
H'_{\text{int}} = \underset{\mathbf{x},i}{\sum}\left(\xi\left(\mathbf{x},i\right)\epsilon\left(\mathbf{x},i\right)\eta_m^{\dagger}\left(\mathbf{x}\right)U_{mn}\left(\mathbf{x},i\right)\eta_n\left(\mathbf{x}+\mathbf{e}_i\right) + h.c\right)
\end{equation}
where, as in the $U(1)$ case, $\xi_{\mathfrak{p}}\left(\mathbf{x},i,j\right),\xi\left(\mathbf{x},i\right)$ are local functions of electric fields, belonging to neighboring links. Furthermore, the electric fields that appear in these phase factors are only those of the $U(1)$ subgroup that completes $SU(N)$ to $U(N)$,
\begin{equation}
E\left(\mathbf{x},i\right)=\sqrt{2N}L_0\left(\mathbf{x},i\right)=\sqrt{2N}R_0\left(\mathbf{x},i\right)
\label{Edef}
\end{equation}
 \cite{Zohar2018b}.

The method we employ is very similar to what we did in the $U(1)$ case. We would like to define $N$ commuting operators $G_m\left(\mathbf{x}\right)$ per vertex, that will satisfy $N$ Gauss laws
\begin{equation}
G_m\left(\mathbf{x}\right)\left|\psi\right\rangle = n_m\left(\mathbf{x}\right)\left|\psi\right\rangle
\label{Gm}
\end{equation}
These will allow us to eliminate each component of the matter spinors independently of the others. In order to do that, consider the $N$ commuting charges of the Cartan subalgebra,
\begin{equation}
Q_{\mu}\left(\mathbf{x}\right)=\Lambda_{\mu m}n_m\left(\mathbf{x}\right)-\sqrt{\frac{N}{2}}\delta_{\mu 0}s\left(\mathbf{x}\right)
\label{Qmu}
\end{equation}
where
\begin{equation}
\Lambda_{\mu m} = \left(T_{\mu}\right)_{mm} \text{    (no summation.)}
\end{equation}
(where $\mu=0,...,N-1,m=1,...,N$).
The normalization of the generators (\ref{trTT}) implies that $\left(\Lambda \Lambda^T\right)_{\mu \nu} = \delta_{\mu \nu}/2$ and therefore $\Lambda^{-1}=2\Lambda^T$.
We now consider the $N$ commuting Gauss laws of the Cartan subalgebra. Using (\ref{GNA}) and (\ref{Qmu}), they can be rewritten as
\begin{equation}
D_{\mu}\left(\mathbf{x}\right)\left|\psi\right\rangle=\left(\Lambda_{\mu m}n_m\left(\mathbf{x}\right)-\sqrt{\frac{N}{2}}\delta_{\mu 0}s\left(\mathbf{x}\right)\right)\left|\psi\right\rangle
\label{Dmu}
\end{equation}
Using $\Lambda^{-1}=2\Lambda^T$ with the above equation, we obtain that Eq. (\ref{Gm}) is satisfied, with
\begin{equation}
G_m\left(\mathbf{x}\right)=2\Lambda_{\mu m}D_{\mu}\left(\mathbf{x}\right)+s\left(\mathbf{x}\right)
\label{Gmdef}
\end{equation}
analogously to (\ref{Gdef}).

Combining (\ref{Gmdef}) with (\ref{ULR}), we obtain the very simple Abelian transformation rules
\begin{equation}
\begin{aligned}
&\left[G_k\left(\mathbf{x}\right),U_{mn}\left(\mathbf{y},i\right)\right]=
\delta\left(\mathbf{x},\mathbf{y}\right)\delta_{km}U_{mn}\left(\mathbf{y},i\right)
\\&-
\delta\left(\mathbf{x},\mathbf{y}+\mathbf{e}_i\right)\delta_{kn}U_{mn}\left(\mathbf{y},i\right)
\text{ (no summation).}
\end{aligned}
\label{UGab}
\end{equation}
This, along with
\begin{equation}
G_m\left(\mathbf{x}\right)\left(G_m\left(\mathbf{x}\right)-1\right)\left|\psi\right\rangle=1
\end{equation}
that follows directly from (\ref{Gm}) as in the Abelian case, allows us to write $H'_{\text{int}}$ projected to the physical Hilbert space,
\begin{widetext}
	\begin{equation}
	H'^{\text{phys}}_{\text{int}} =  \underset{\mathbf{x},i,m,n}{\sum}\left(\xi\left(\mathbf{x},i\right)\epsilon\left(\mathbf{x},i\right)P^m_1\left(\mathbf{x}\right)\eta_m^{\dagger}\left(\mathbf{x}\right)U_{mn}\left(\mathbf{x},i\right)\eta_n\left(\mathbf{x}+\mathbf{e}_i\right)P^n_0\left(\mathbf{x}+\mathbf{e}_i\right) + h.c\right)
	\end{equation}
where $P^{m}_g\left(\mathbf{x}\right)$ projects $G_m\left(\mathbf{x}\right)$ to $g$. The Gauss laws also enable us to rewrite, within the physical Hilbert space, the mass part of the Hamiltonian as
	\begin{equation}
	H^{\text{phys}}_M = M\underset{\mathbf{x}}{\sum}\left(-1\right)^{s\left(\mathbf{x}\right)}G\left(\mathbf{x}\right) = 2M\underset{\mathbf{x},i}{\sum}\left(-1\right)^{s\left(\mathbf{x}\right)}
	E\left(\mathbf{x},i\right) + \text{const.}
	\end{equation}
\end{widetext}
where $E$ is the Abelian, $U(1) \nsubseteq SU(N)$ electric field (\ref{Edef}).

With all that in hand, we can finally construct the local building blocks of the transformation that removes the matter. On each vertex, we define the unitaries
\begin{equation}
u_m\left(\mathbf{x}\right) = \eta_m\left(\mathbf{x}\right)+\eta^{\dagger}_m\left(\mathbf{x}\right)
\end{equation}
They do not commute with each other on-site, but do commute on different sites, which allows us to construct the transformation from local, commuting pieces. We will have then a transformation $\mathcal{U}=\underset{\mathbf{x}}{\prod}\mathcal{U}\left(\mathbf{x}\right)$ with the local, commuting transformations
\begin{equation}
\mathcal{U}\left(\mathbf{x}\right) = u_N\left(\mathbf{x}\right)^{G_N\left(\mathbf{x}\right)} \cdots u_1\left(\mathbf{x}\right)^{G_1\left(\mathbf{x}\right)}
\end{equation}
The order of the $u_m^{G_m}$ in the product matters, since they do not commute, but different orders give rise, finally, to physically equivalent results, and as the local products commute with one another it does not matter. Eventually, one obtains that
\begin{widetext}
\begin{equation}
\begin{aligned}
&\mathcal{U} P^m_1\left(\mathbf{x}\right)\eta_m^{\dagger}\left(\mathbf{x}\right)U_{mn}\left(\mathbf{x},i\right)\eta_n\left(\mathbf{x}+\mathbf{e}_i\right)P^n_0\left(\mathbf{x}+\mathbf{e}_i\right)
\mathcal{U}^{\dagger} =\\
&\left(-1\right)^{\overset{m-1}{\underset{i=1}{\sum}}G_i\left(\mathbf{x}\right)}
P^m_1\left(\mathbf{x}\right)\eta_m\left(\mathbf{x}\right)\eta_m^{\dagger}\left(\mathbf{x}\right)U_{mn}\left(\mathbf{x},i\right)
\eta_n\left(\mathbf{x}+\mathbf{e}_i\right)\eta^{\dagger}_n\left(\mathbf{x}+\mathbf{e}_i\right)P^n_0\left(\mathbf{x}+\mathbf{e}_i\right)
\left(-1\right)^{\overset{n-1}{\underset{i=1}{\sum}}G_i\left(\mathbf{x}+\mathbf{e}_i\right)}
\end{aligned}
\end{equation}

One can see that the matter is completely decoupled here: the only instances of which is through projectors to $n_m=0$ everywhere. Therefore, we conclude that the original Hamiltonian is equivalent to this of a matter-less theory without local symmetries (all the Gauss laws are transformed to trivial $0=0$ equations as in the Abelian case), whose Hamiltonian is
\begin{equation}
\begin{aligned}
&\underset{\mathbf{x}}{\otimes}\left\langle n_m=0\right|_{\mathbf{x}}
\mathcal{U}H'^{\text{phys}}\mathcal{U}^{\dagger}
\underset{\mathbf{x}}{\otimes}\left|n_m=0\right\rangle_{\mathbf{x}}
=
H'_{KS} + 2M\underset{\mathbf{x},i}{\sum}\left(-1\right)^{s\left(\mathbf{x}\right)}
E\left(\mathbf{x},i\right)
\\&+
\underset{\mathbf{x},i,m,n}{\sum}\left(\epsilon\left(\mathbf{x},i\right)\xi\left(\mathbf{x},i\right)\left(-1\right)^{\overset{m-1}{\underset{i=1}{\sum}}G_i\left(\mathbf{x}\right)}
P^m_1\left(\mathbf{x}\right)U_{mn}\left(\mathbf{x},i\right)
P^n_0\left(\mathbf{x}+\mathbf{e}_i\right)
\left(-1\right)^{\overset{n-1}{\underset{i=1}{\sum}}G_i\left(\mathbf{x}+\mathbf{e}_i\right)}+ h.c\right)
\end{aligned}
\end{equation}
\end{widetext}

The price we pay for having originally fermionic matter is double: the factors $\xi\left(\mathbf{x},i\right)\left(-1\right)^{\overset{m-1}{\underset{i=1}{\sum}}G_i\left(\mathbf{x}\right)}\left(-1\right)^{\overset{n-1}{\underset{i=1}{\sum}}G_i\left(\mathbf{x}+\mathbf{e}_i\right)}$
for the statistics, and the projectors $P^m_1\left(\mathbf{x}\right)$ and $P^n_0\left(\mathbf{x}+\mathbf{e}_i\right)$ for the finiteness of the local matter spaces. Both extend the interaction range to nearest neighbor links, but not beyond.

What happens if we wish to include a larger representation of the fermions? In this case, the method will not work, for the simple reason that we have exactly $N$ commuting Gauss laws in the Cartan subalgebra, that can be inverted to define $N$ different $G_m$ operators. These can only correspond to $N$ fermionic number operators per vertex - no more. Once again, we will also not be able to extend the method for non-staggered, or flavored fermions.

\subsection{The $SU(N)$ case}
Finally, we wish to discuss the case of another very relevant gauge group - $SU(N)$. Can we repeat the same procedure there? First, of course, we need to map fermionic matter to hard-core bosonic ones. According to \cite{Zohar2018b}, this is possible only for $SU(2N)$ without extra ingredients. If one wishes to do it for $SU(2N+1)$, an auxiliary $\mathbb{Z}_2$ gauge field has to be introduced, without dynamics, to absorb the parity of the fermions and enable the transformation.

However, even if we do that, it will not allow us to repeat the procedure used for $U(N)$, since in $SU(N)$ we only have $N-1$ generators in the Cartan subalgebra - and the fundamental representation is $N$ dimensional. We only have $N-1$ Gauss laws of the form (\ref{Dmu}) per vertex, while we need $N$ equations of the form (\ref{Gmdef}). The inversion discussed in the $U(N)$ case will not be possible now (the matrix $\Lambda$ is no longer square). As we shall show, for $SU(2N)$ there is only one way to proceed, through the introduction of an auxiliary $U(1)$ gauge field, but for $SU(2N+1)$ one can also use another method, the auxiliary  $\mathbb{Z}_2$ gauge field introduced in \cite{Zohar2018b} is enough.

\subsubsection{The $SU(2N+1)$ case}
We seek for another, independent equation, that will add up to the $N-1$ Gauss laws (\ref{Dmu}), allowing us to explicitly solve for each $n_m$ separately. One could try to think of using the center $\mathbb{Z}_N$ symmetry of $SU(N)$ - an Abelian subgroup that is a subgroup of the missing $U(1)$ component we had in $U(N)$. However, the elements of the center are diagonal $SU(N)$ elements that are generated with the $N-1$ Cartan generators, so this will introduce no further, independent equation.

Suppose that we add an auxiliary $\mathbb{Z}_2$ field, following the procedure of \cite{Zohar2018b}. While for $SU(2N+1)$ it is required for the conversion to hard-core bosons, it is not required for $SU(2N)$ - but we can try to add it nevertheless in both cases. Then, on each link we introduce an extra $\mathbb{Z}_2$ Hilbert space - a two level space of a single spin - and wherever $U_{mn}\left(\mathbf{x},i\right)$ appears in the Hamiltonian, we replace it by $U'_{mn}\left(\mathbf{x},i\right)=U_{mn}\left(\mathbf{x},i\right)Z\left(\mathbf{x},i\right)$, where $Z\left(\mathbf{x},i\right)$ is the Pauli z operator acting on the auxiliary field on that link. We do not include dynamics for this field, as explained in \cite{Zohar2018b}.

The auxiliary field, when coupled to the matter in the above way, introduces, in the extended Hilbert space, an extra local $\mathbb{Z}_2$, given by
\begin{equation}
X\left(\mathbf{x}\right)\left|\psi\right\rangle
\equiv
\underset{i}{\prod}\left[X\left(\mathbf{x},i\right)X\left(\mathbf{x}-\mathbf{e}_i,i\right)\right]\left|\psi\right\rangle = \left(-1\right)^{\underset{m}{\sum}n_m\left(\mathbf{x}\right)}\left|\psi\right\rangle
\label{Z2}
\end{equation}
 both before and after transforming to hard-core bosons.

Instead of the Cartan-Weyl basis we used in the $U(N)$ case, we will now use another form for the traceless generators, replacing $\Lambda$ by
\begin{equation}
\Lambda'_{\mu m} = \delta_{\mu m}-\delta_{N m}; \quad \quad \mu=1,...,N-1,m=1,...,m
\end{equation}
This brings the $N-1$ Cartan Gauss laws (\ref{Dmu}) to the form
\begin{equation}
D_{\mu} \left(\mathbf{x}\right) \left|\psi\right\rangle= \left(n_{\mu}\left(\mathbf{x}\right)-n_{N}\left(\mathbf{x}\right)\right)\left|\psi\right\rangle
\label{Dmus}
\end{equation}
Note that
\begin{equation}
D\left(\mathbf{x}\right)\left|\psi\right\rangle \equiv \underset{\mu}{\sum}D_{\mu}\left(\mathbf{x}\right)\left|\psi\right\rangle=\left(\underset{m}{\sum}n_m\left(\mathbf{x}\right)-Nn_N\left(\mathbf{x}\right)\right)\left|\psi\right\rangle
\end{equation}
and in particular
\begin{equation}
\left(-1\right)^{D\left(\mathbf{x}\right)}\left|\psi\right\rangle =
\left(-1\right)^{Nn_N\left(\mathbf{x}\right)}
\left(-1\right)^{\underset{m}{\sum}n_m\left(\mathbf{x}\right)}
\left|\psi\right\rangle
\end{equation}
Using the auxiliary $\mathbb{Z}_2$ local symmetry we obtain
\begin{equation}
\left(-1\right)^{D\left(\mathbf{x}\right)}\left|\psi\right\rangle =
\left(-1\right)^{Nn_N\left(\mathbf{x}\right)}
X\left(\mathbf{x}\right)\left|\psi\right\rangle
\end{equation}
and now the roads for even and odd values of $N$ split. In the even case, $\left(-1\right)^{Nn_N\left(\mathbf{x}\right)}=1$, and the above equation shows us that we gain nothing from introducing the auxiliary field, since a local $\mathbb{Z}_2$ exists in the $SU(2N)$ case anyway, as it is a subgroup of the group's center, $\mathbb{Z}_{2N}$. Therefore, the current discussion can only be valid for $SU(2N+1)$, where, since $\left(-1\right)^{Nn_N\left(\mathbf{x}\right)}=\left(-1\right)^{n_N\left(\mathbf{x}\right)}=1-2n_N\left(\mathbf{x}\right)$,
we obtain the desired equation
\begin{equation}
n_N\left(\mathbf{x}\right)\left|\psi\right\rangle = \frac{1}{2}\left(1-X\left(\mathbf{x}\right)\left(-1\right)^{D\left(\mathbf{x}\right)}\right)\left|\psi\right\rangle
\end{equation}
and immediately define
\begin{equation}
G_N\left(\mathbf{x}\right) = \frac{1}{2}\left(1-X\left(\mathbf{x}\right)\left(-1\right)^{D\left(\mathbf{x}\right)}\right)
\label{GNs}
\end{equation}
Combining it with (\ref{Dmus}), we can finally define
\begin{equation}
G_{\mu}\left(\mathbf{x}\right) = D_{\mu}\left(\mathbf{x}\right) + G_N\left(\mathbf{x}\right)
\end{equation}
as well, which completes a linearly independent set of $N$ mutually commuting operators $G_m\left(\mathbf{x}\right)$ satisfying
$G_m\left(\mathbf{x}\right)\left|\psi\right\rangle = n_m\left(\mathbf{x}\right)\left|\psi\right\rangle$ as in the $U(N)$ case - but constructed differently. One can then construct the desired transformation $\mathcal{U}$ using these $G_m\left(\mathbf{x}\right)$ operators.

As before, one needs to constrain the operators $G_{\mu}$ to have only $0,1$ eigenvalues, which gives rise to local constraints in the final, transformed model. However, note that such a constraint is not required for $G_N$ now, since this operator only has in its spectrum $0,1$ anyway.

For the $SU(2N)$ case, however, we will have to use another method, that will be discussed next.

\subsubsection{The $SU(2N)$ case}
In order to solve the $SU(2N)$ case - where we simply do not have enough commuting, linearly independent equations to invert - we will need to introduce an auxiliary $U(1)$ gauge field, and embed $SU(N)$ in $U(N)$. This applies to any $N$, and thus we will not restrict ourselves only to $SU(2N)$ in the discussion. As $U(N)$ satisfies the requirements for transforming fermions to hard-core bosons, no $\mathbb{Z}_2$ field has to be introduced, and the $U(1)$ auxiliary field should be introduced before converting the fermions to bosons.

On each link of our $SU(N)$ system we introduce an additional $U(1)$ gauge field, with electric field $E$ and phase operator $\phi$. We define the extended $U(N)$ group element operators,
\begin{equation}
U'\left(\mathbf{x},i\right)=U\left(\mathbf{x},i\right)e^{i\phi\left(\mathbf{x},i\right)}
\end{equation}
that add the new $U(1)$ component to the former $U$ operators. We modify the Hamiltonian $H_{SU(N)}$ to another one, $H_{U(N)}$, by replacing any $SU(N)$ operator $U$ by the extended $U(N)$ operator $U'$, without adding any dynamics: the $\mathbf{J}^2$ terms in $H_{KS}$ are left only with the $SU(N)$ generators. Then, the original $SU(N)$ Hamiltonian may be obtained by projecting the new $U(N)$ Hamiltonian to a configuration with $\phi=0$ everywhere: since $\left\langle \phi=0\right|U'\left|\phi=0\right\rangle = U$, and $H_{U(N)}$ completely commutes with all the $\phi$ operators, we get that
\begin{equation}
\left\langle\left\{\phi=0\right\}\right|H_{U(N)}\left|\left\{\phi=0\right\}\right\rangle = H_{SU(N)}
\end{equation}
Therefore, if $\left|\psi\right\rangle$ is an eigenstate of $H_{SU(N)}$, $\left|\psi\right\rangle \otimes \left|\left\{\phi=0\right\}\right\rangle$ is an eigenstate of the extended $H_{U(N)}$ with the same energy. However, this state is not invariant under the complete set of $U(N)$ gauge transformations, since the Abelian phase is fixed on all the links.

We therefore define, for each state $\left|\psi\right\rangle$ in the $SU(N)$ physical Hilbert space, the state $\left|\Psi\right\rangle$ as follows:
\begin{equation}
\left|\Psi\right\rangle = \mathcal{N}^{-1/2}\int \mathcal{D}\alpha e^{-i\sqrt{2N}\underset{\mathbf{x}}{\sum}\left(D_0\left(\mathbf{x}\right)-Q_0\left(\mathbf{x}\right)\right)\alpha\left(\mathbf{x}\right)}\left|\psi\right\rangle \otimes \left|\left\{\phi=0\right\}\right\rangle
\label{Psi}
\end{equation}
where $\mathcal{D}\alpha \equiv \underset{\mathbf{x}}{\prod}d\alpha\left(\mathbf{x}\right) $, $\left\{\alpha\left(\mathbf{x}\right)\right\}$ is a set of local phases, and $\mathcal{N}$ is a normalization constant.
The state $\left|\psi\right\rangle$ satisfies the $N^2-1$ Gauss laws (\ref{GNA}) corresponding to $SU(N)$. The state $\left|\Psi\right\rangle$ constructed from it using (\ref{Psi}) preserves this symmetry, and adds up the missing $U(N)$ Gauss law as well - it has the complete $U(N)$ gauge invariance.

Let us show that the mapping $\left|\psi\right\rangle \rightarrow \left|\Psi\right\rangle$ is an isomorphism.
We begin with the norm of $\left|\Psi\right\rangle$.
\begin{widetext}
\begin{equation}
\left\langle\Psi|\Psi\right\rangle = \mathcal{N}^{-1}\int \mathcal{D}\alpha \mathcal{D}\beta \left\langle \psi \right|
e^{i\sqrt{2N}\underset{\mathbf{x}}{\sum} Q\left(\mathbf{x}\right)\left(\alpha\left(\mathbf{x}\right)-\beta\left(\mathbf{x}\right)\right)} \left|\psi\right\rangle
\left\langle \left\{\phi=0\right\} \right|
e^{-i\sqrt{2N}\underset{\mathbf{x}}{\sum} D_0\left(\mathbf{x}\right)\left(\alpha\left(\mathbf{x}\right)-\beta\left(\mathbf{x}\right)\right)}
\left|\left\{\phi=0\right\}\right\rangle
\end{equation}
Changing variables of integration, one obtains
\begin{equation}
\left\langle\Psi|\Psi\right\rangle = \mathcal{N}^{-1} \left(\int\mathcal{D}\beta\right) \int\mathcal{D}\alpha \left\langle \psi \right|
e^{i\sqrt{2N}\underset{\mathbf{x}}{\sum} Q\left(\mathbf{x}\right)\alpha\left(\mathbf{x}\right)} \left|\psi\right\rangle
\left\langle \left\{\phi=0\right\} \right|
e^{-i\sqrt{2N}\underset{\mathbf{x}}{\sum} D_0\left(\mathbf{x}\right)\alpha\left(\mathbf{x}\right)}
\left|\left\{\phi=0\right\}\right\rangle
\end{equation}
Since $e^{-i\sqrt{2N}\underset{\mathbf{x}}{\sum} D_0\left(\mathbf{x}\right)\alpha\left(\mathbf{x}\right)}$ generates a pure-gauge transformation on the $U(1)$ part, we obtain that
\begin{equation}
\left\langle \left\{\phi=0\right\} \right|
e^{-i\sqrt{2N}\underset{\mathbf{x}}{\sum} D_0\left(\mathbf{x}\right)\alpha\left(\mathbf{x}\right)}
\left|\left\{\phi=0\right\}\right\rangle
=\underset{\mathbf{x},i}{\prod}
\left\langle \phi=0 |
\phi=\alpha\left(\mathbf{x}\right)-\alpha\left(\mathbf{x}+\mathbf{e}_i\right)\right\rangle
=\underset{\mathbf{x},i}{\prod}\delta\left(\alpha\left(\mathbf{x}\right)-\alpha\left(\mathbf{x}+\mathbf{e}_i\right)\right)
\end{equation}

We have more delta functions than integrations: the number of delta functions is the number of links, $\mathcal{N}_L$, and the number of integrations is the number of vertices, $\mathcal{N}_V$ (both depend on the system size and topology). We choose a path of $\mathcal{N}_V$ links that goes through all the vertices and use it to perform all the integrations, and then we are left with the same phase everywhere - $\alpha_0$, and an infinite constant, $\delta\left(0\right)^{\mathcal{N}_L-\mathcal{N}_V}$. Finally we obtain for the norm
\begin{equation}
\left\langle\Psi|\Psi\right\rangle = \mathcal{N}^{-1}\delta\left(0\right)^{\mathcal{N}_L-\mathcal{N}_V} \left(2\pi\right)^{\mathcal{N}_V} \left\langle \psi \right|
e^{i\sqrt{2N}\left(\underset{\mathbf{x}}{\sum} Q\left(\mathbf{x}\right)\right)\alpha_0} \left|\psi\right\rangle
\end{equation}
The state $\left|\psi\right\rangle$ does not have a local $U(1)$ symmetry, but has a global one (conservation of total number of particles). The transformation we are left with, for the computation is the norm, is global, and hence $\left\langle \psi \right|
e^{i\sqrt{2N}\left(\underset{\mathbf{x}}{\sum} Q\left(\mathbf{x}\right)\right)\alpha_0} \left|\psi\right\rangle = \left\langle \psi |\psi\right\rangle$. So if we set
\begin{equation}
\mathcal{N} = \delta\left(0\right)^{\mathcal{N}_L-\mathcal{N}_V} \left(2\pi\right)^{\mathcal{N}_V}
\end{equation}
we get that the mapping $\left|\psi\right\rangle \rightarrow \left|\Psi\right\rangle$ preserves the norm.

What about inner products? Using exactly the same arguments as above, we obtain that, in general,
\begin{equation}
\left\langle \Psi_1 | \Psi_2\right\rangle = \left\langle \psi_1 | \psi_2\right\rangle
\end{equation}

Next, we show that it preserves the Hamiltonian matrix elements.
\begin{equation}
\left\langle \Psi_1 \right| H_{U(N)} \left| \Psi_2\right\rangle = \int \mathcal{D}\alpha \mathcal{D}\beta \mathcal{N}^{-1}\left\langle\psi_1\right| \otimes \left\langle \left\{\phi=0\right\}\right|
e^{i\sqrt{2N}\underset{\mathbf{x}}{\sum}\left(D_0\left(\mathbf{x}\right)-Q_0\left(\mathbf{x}\right)\right)\beta\left(\mathbf{x}\right)}
H_{U(N)}
e^{-i\sqrt{2N}\underset{\mathbf{x}}{\sum}\left(D_0\left(\mathbf{x}\right)-Q_0\left(\mathbf{x}\right)\right)\alpha\left(\mathbf{x}\right)}
\left|\psi_2\right\rangle \otimes \left| \left\{\phi=0\right\}\right\rangle
\end{equation}
But since $e^{-i\sqrt{2N}\underset{\mathbf{x}}{\sum}\left(D_0\left(\mathbf{x}\right)-Q_0\left(\mathbf{x}\right)\right)\alpha\left(\mathbf{x}\right)}$ is a gauge transformation it commutes with $H_{U(N)}$, and $H_{U(N)}\left|\psi_2\right\rangle \otimes \left| \left\{\phi=0\right\}\right\rangle = H_{U(N)}$, and $H_{U(N)}\left|\psi_2\right\rangle \otimes \left| \left\{\phi=0\right\}\right\rangle = \left(H_{SU(N)}\left|\psi_2\right\rangle\right) \otimes \left| \left\{\phi=0\right\}\right\rangle$.
Therefore, and using the same methods for the integration etc.,
\begin{equation}
\left\langle \Psi_1 \right| H_{U(N)} \left| \Psi_2\right\rangle = \left\langle \psi_1 \right| H_{SU(N)} \left| \psi_2\right\rangle
\end{equation}
\end{widetext}

Indeed, the physical Hilbert spaces of $H_{SU(N)}$ and $H_{U(N)}$ are isomorphic, and one may use $H_{U(N)}$ to study $H_{SU(N)}$. This is true for open boundary conditions, where there are no topological sectors. In the case of periodic boundaries, only states in the same topological sector are connected unitarily to $\left| \left\{\phi=0\right\}\right\rangle$, and hence the right sector must be sought. But in general we see that studying the extended theory where $SU(N)$ is embedded into $U(N)$ is physically equivalent. Therefore, one can use the above prescription to embed an $SU(N)$ lattice gauge theory into a $U(N)$ one, and then eliminate the fermions as explained in the previous section.

\section{Summary and Conclusions}
In this paper we have shown how to completely remove staggered fundamental fermionic matter from $U(N)$ and $SU(N)$ lattice gauge theories, by solving the Gauss law for the matter, which, unlike the solution for the electric field, is independent of the dimension. We extended the well known procedure for complex scalar fields, arising from the Higgs mechanism, to fermionic matter and showed that it can be eliminated as well, by making use of the gauge symmetry and breaking it. However, unlike in the scalar case, when the matter is fermionic one must introduce other local constraints, accounting for the finiteness of the local Hilbert spaces of the original matter, and slightly extend the range of interactions due to the same fact, and also accounting for the fermionic statistics, building up on the result of \cite{Zohar2018b}.

This opens the way to possibly easier variational and numerical computations for Hamiltonian lattice gauge theories, without having to deal with extra degrees of freedom for the matter - in particular with fermionic Fock spaces, as well as for quantum simulation of lattice gauge theories \cite{Tagliacozzo2013,Wiese2013,Zohar2015a,Dalmonte2016,Preskill2018} with fermionic matter, without the actual use of fermionic degrees of freedom in the simulator.

After the completion of this work, we became aware of another work \cite{Surace2019} discussing the elimination of the matter degrees of freedom, in 1+1 compact QED (the Schwinger model).

\begin{acknowledgments}
 EZ would like to thank  David B. Kaplan, Martin J. Savage and John Preskill for insightful discussions. JIC is partially supported by the EU, ERC grant QUENOCOBA 742102. This work was supported by the EU-QUANTERA project QTFLAG (BMBF grant No. 13N14780).
  \end{acknowledgments}

\bibliography{ref}
\end{document}